# Development and prospect of Very Small Angle Neutron Scattering (VSANS) Techniques

Taisen Zuo(左太森)[1,2,3], He Cheng(程贺)[1,2;1], Yuan-Bo Chen(陈元柏)[1,2;2], Fang-Wei Wang(王芳卫)[2,4]

[1] China Spallation Neutron Source (CSNS), Institute of High Energy Physics (IHEP), Chinese Academy of Sciences (CAS), Dongguan 523803, China

[2] Dongguan Institute of Neutron Science (DINS), Dongguan 523808, China

[3] University of Chinese Academy of Sciences, Beijing 100049, China

[4] Institute of Physics, Chinese Academy of Sciences, Beijing 100190, China

**Abstract:** Very Small Angle Neutron Scattering (VSANS) is an upgrade of the traditional Small Angle Neutron Scattering (SANS) technique which can cover three orders of magnitude of length scale from one nanometer to one micrometer. It is a powerful tool for structure calibration in polymer science, biology, material science and condensed matter physics. Since the first VSANS instrument, D11 in Grenoble, was built in 1972, new collimation techniques, focusing optics (multi-beam converging apertures, material or magnetic lenses, and focusing mirrors) and higher resolution detectors combined with the long flight paths and long incident neutron wavelengths have been developed. In this paper, a detailed review is given of the development, principles and application conditions of various VSANS techniques. Then, beam current gain factors are calculated to evaluate those techniques. A VSANS design for the China Spallation Neutron Source (CSNS) is thereby presented.

## 1 Introduction

Since its discovery by Chadwick in 1932, the neutron has been widely used to exploit the structure and dynamics in materials at various energy and length scales. The Nobel Prize in physics 1995 was awarded to Clifford G. Shull "for the development of the neutron diffraction technique" and Bertram N. Brockhouse "for the development of neutron spectroscopy" [1]. Their pioneer work in the 1940s and 1950s help to answer questions about where atoms "are" and what atoms "do". With the development of modern science and technology, neutron instrumentation and related methodology have diversified and become specialized. Methods and instrumentations such as Small Angle Neutron Scattering (SANS) have been developed to characterize large scale structures in soft and hard condensed matter physics.

Theory and technology developed in Small-Angle Light Scattering (SLS) and Small-Angle X-ray scattering (SAXS) by Guiniur, Deby and Porod were introduced to Small Angle Neutron Scattering (SANS) in the 1970s and 1980s [2, 3]. Neutron source centers with SANS instruments were then built in Jülich [4] and Institute Laue-Langevin (ILL) Grenoble [5]. The unique properties of the neutron make it a complementary technique to X-ray scattering. First, the neutron interacts with the nucleus rather than with electron clouds, so it is sensitive to light elements and isotopes (complementary to X-ray, which is sensitive to elements with high molar mass). Second, the small scattering cross-section of most elements makes it easy for neutrons to penetrate centimeters into materials, thus complicated sample environment and time resolved observations are possible in neutron scattering experiments. Third, thermal and cold neutrons bear much less energy than X-rays with the same wavelength, so there are no heat

∗ Supported by National Natural Science Foundation of China (21474119) and (11305191)
1) E-mail:chenghe@ihep.ac.cn
2) E-mail: chenyb@ ihep.ac.cn

and decomposition effects on the sample, which is very important in biology.

In practice, no single neutron instrument can cover all the length scales of interest. For example, biology macromolecules usually have multi-scale structures ranging from angstroms to tens of micrometers [6]. In order to "see" structures over such a wide range, instruments with different scattering vector (Q) ranges must be used, i.e., neutron diffraction with Q range from 0.1 Å$^{-1}$ to 20 Å$^{-1}$, small angle neutron scattering with Q range from 0.004 Å$^{-1}$ to 0.6 Å$^{-1}$ and ultra-small-angle neutron scattering with Q range from 0.00005 Å$^{-1}$ to 0.01 Å$^{-1}$ must be combined. However, different instruments have unique sample environments, specific instrument resolutions and backgrounds, resulting in difficulties in combining data from different sources. Data analysis is another challenge. The users have to write software which contains the resolution corrections of each instrument. Moreover, neutron beam time is precious. It is almost impossible for any user to get beam time on three different instruments in a half-year cycle. Therefore, traditional small angle neutron scattering instruments, which cover the middle Q range from 0.004 to 0.6 Å$^{-1}$, must be extended to both sides of the Q range to satisfy the requirements of most users.

There are two tendencies in the upgrade of traditional small angle neutron scattering instruments. The first is to the high Q range. A typical example is NIMROD in Rutherford Appleton Laboratory (ISIS) [7], which combines the advantage of SANS and diffractometer and covers the Q range from 0.01 Å$^{-1}$ to 20 Å$^{-1}$. The second tendency goes to low Q range or larger length scale. D11 in ILL is the first such example, combining the advantages of both SANS and USANS, and measuring sizes from 1 nanometer to 1 micrometer. In this paper, we will concentrate on the development and perspective of the latter VSANS techniques.

At the very beginning, those SANS instruments, i.e., Loq [8] (ISIS), and 8m SANS [9] (National Bureau of Standard/National Institute of Standards and Technology NBS/NIST) were not equipped with any reflective or refractive focusing devices. They simply used two separated pinholes to collimate the incident beam. So, the first generation of SANS was also called pinhole SANS. Minimum Q accessible for pinhole SANS cannot be lower than 0.003 A$^{-1}$. The use of a series of MgF$_2$ lenses in a pinhole system successfully decreases the minimum Q to 0.001 Å$^{-1}$. Because this kind of material is easy to handle and operate, pinhole SANS instruments with optic lenses are widely used in almost all of the reactor based neutron centers nowadays [10-12], and can be regarded as the second generation of SANS instruments. Unfortunately, they can only measure length scales smaller than 300 nm, because in the Guinier region, $QR_g$ should be smaller than 1, where $R_g$ is the radius of gyration. To access smaller Q, lots of effort has been made and progress has been achieved; for example, an extremely long instrument with long wavelength neutrons at D11 in ILL [13], and multi-slit and multi-pinhole VSANS in Laboratoire Leon Brillouin (LLB) [14]. Some other VSANS are still under development. T. Oku et al. tried to use magnetic lenses to focus neutrons [15] [16]. D. Liu et al. worked on a toroidal Wolter mirror to focus neutrons [17] [18].

Based on these previous works, we calculated the gain factors of the different techniques used to assist our decision in designing a VSANS instrument for the China Spallation Neutron Source (CSNS).

## 2 Experimental method

Virtual experiments or simulations were carried out with McStacs2.2 [19]. Source data of the coupled hydrogen moderator used in the virtual experiments was provided by the neutronics group of CSNS [20], and SNS_source COMPONENT was used as the generator of virtual neutrons.

## 3 Introduction to possible VSANS techniques

In order to get to smaller scattering vectors, VSANS requires longer collimation and precise neutron focusing techniques. In the following, the development of VSANS theory and instrumentation will be presented. To simplify the deduction process, ideal pinhole geometry will be used as shown in Fig. 1a, where Source-to-Sample Distance is assumed to be the same as the Sample-to-Detector Distance, $L_{SSD}=L_{SDD}=L$, and the radius of the source aperture $R_{1P}$ is twice as big as that of the sample aperture $R_{2P}$; gravity effect, parasite scattering and spatial resolution of the detector are all omitted. Then the beam current gain factors of each VSANS technique over ideal pinhole geometry is calculated and compared. Based on the analysis, the design of a future VSANS instrument at CSNS is introduced. Note that Fig. 1-4 are only schematic illustrations of different techniques, and are used to give an intuitive description of the calculation of gain factors.

## 3.1 Extreme long instrument D11

The first VSANS instrument D11[5] was built in ILL, France in 1972. It is an extremely long instrument with pinhole geometry. With a total length of 80 m, the minimum Q can be $5\times10^{-4}$ Å$^{-1}$ [21]. D11 has been a world-leading instrument for three decades and it is still considered a benchmark for VSANS [13]. After a major upgrade in 2008-2009, its lowest published Q is $3.4\times10^{-4}$ Å$^{-1}$ [13] with incident neutrons at 20 Å. By countering the effect of gravity with a reflecting mirror and focusing with two MgF$_2$ lenses, 23 Å neutrons can be used to further lower the Q$_{min}$ [22].

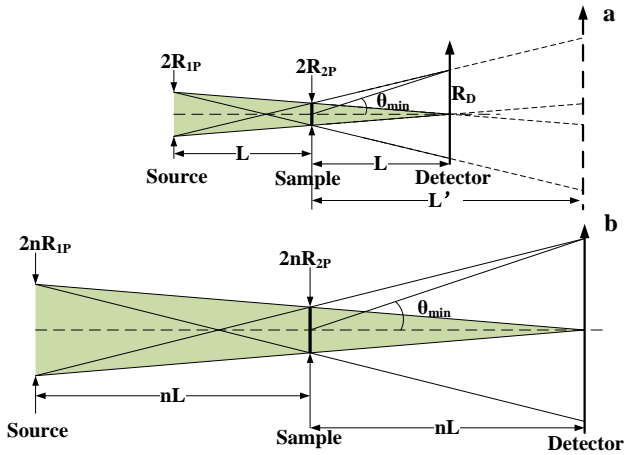

Fig. 1 a. Pinhole geometry; b. Extremely long pinhole geometry (Subscript 1 denotes source aperutre, 2 denotes sample aperutre, P denotes Pinhole, E denotes Extremely long, respectively).

As shown in Fig. 1(a), the collimation system of pinhole geometry is composed of two circular apertures separated by a distance of L. Optimal geometry for the scattering vector can be achieved if sample to detector distance L' equals L and sample aperture diameter 2R$_{2P}$ equals the radius of the source aperture radius R$_{1P}$ [23], and the two apertures form a cone with its vertex at the surface of the detector [24]. The minimum accessible angle for this geometry is $\theta_{min}=R_D/L$, corresponding to a minimum scattering vector [25]

$$Q_{minP} = \frac{4\pi}{\lambda}\sin\left(\frac{\theta_{min}}{2}\right) = k\theta_{min} = k\frac{R_D}{L} = k\frac{2R_{1P}}{L} \quad (1)$$

where k=2π/λ. In order to understand the potential benefit of extremely long collimation, a comparison of neutron beam current at the same Q$_{min}$ will be given.

Beam current I (neutrons per second or n/s) at the sample site could be rendered as [26]

$$I = \frac{d^2\phi_s}{d\Omega d\lambda}\Delta\lambda\frac{A_1A_2}{L^2} \quad (2)$$

where $\phi_s$ (n/cm$^2$/s) denotes neutrons emitted per unit area of the moderator per second to 2π solid angle, $(d^2\phi_s)/(d\Omega d\lambda)$ is the source brightness, A$_1$=πR$_{1P}$$^2$ is the area of the source aperture, A$_2$=πR$_{2P}$$^2$ is the area of the sample aperture, and Δλ is the wavelength spread in the incident beam. I/A$_2$ (n/cm$^2$/s) is the neutron flux (n/cm$^2$/s) at the sample.

Using Eq. (1) and Eq. (2), beam current at the sample site of traditional pinhole geometry is

$$I_P = \frac{d^2\phi_s}{d\Omega d\lambda}\Delta\lambda\left(\frac{\pi}{4}\right)A_{2P}\left(\frac{Q_{min}}{k}\right)^2 \quad (3)$$

In the same way, beam current at the sample site of extremely long pinhole geometry is

$$I_E = \frac{d^2\phi_s}{d\Omega d\lambda}\Delta\lambda\left(\frac{\pi}{4}\right)A_{2E}\left(\frac{Q_{min}}{k}\right)^2 \quad (4)$$

If Eq. (4) is divided by Eq. (3) and Q$_{min}$ is eliminated, the gain of extremely long pinhole geometry is

$$\text{Gain} = \frac{I_E}{I_P} = \frac{\pi(nR_{2T})^2}{\pi R_{2P}^2} = n^2 \quad (5)$$

Eq. (5) indicates that if the length of the instrument was prolonged n times, the neutron beam current at the same Q$_{min}$ would increase n$^2$ times. At the same time, sample volume increases n$^2$ times, which means all the gain comes from the increase of sample volume. However, the instrument cannot be too long, especially for Time of Flight (TOF) instruments. The longer the instrument, the narrower the neutron wave band and lower the neutron beam current at the sample position. Newly built or proposed VSANS instruments are usually combinations of medium length (L=20 m) and proper focusing techniques [27, 28].

## 3.2 Multi-slit and multi-pinhole apertures

The concepts of multichannel collimator converging diaphragms or soller collimators were proposed in the 1970s by A. C. Nunes and J. M. Carpenter et al. [29, 30]. The first successful testing device of converging multi-pinhole apertures was accomplished in NIST by C. J. Glinka et al. in 1986 [31]. Crossed converging soller collimators were successfully installed in the Intense Pulsed Neutron Source (IPNS) [32]. Other kinds of multichannel collimator techniques have also been introduced thereafter [33-36]. All of those efforts proved that converging diaphragms or apertures are more advantageous over other multichannel techniques, because they minimize collimation materials that may cause parasitic scattering, and their collimation length can be comparatively long. In recent years, new VSANS instruments based on converging multi-slit or multi-pinhole apertures have emerged. In 2006, a compact VSANS with 4 m long collimation using both multi-slit and multi-pinhole apertures was built by S. Desert et al. at LLB, France [14], with minimum Q of $2 \times 10^{-4}$ Å$^{-1}$; In 2006, Helmholtz-Zentrum Berlin (HZB) started to build a new time-of-flight small angle scattering instrument V16/VSANS [37]. In 2009, J. Barker began to build a new VSANS with both multi-slit and multi-pinhole options at NIST [27]. A new SANS instrument, BILBY [38], was built at Australia's research reactor (ANSTO) with multi-slit option and the instrument is now in commissioning phase. In 2014 S. Jaksch et. al. proposed a new VSANS instrument [39] with multi-slit option for the Europe Spallation Source( ESS) [40].

Multichannel focusing collimators do not really focus the neutrons by changing their trajectories. They just collimate multiple beams to the same point of the detector as shown in Fig. 2. For the multi-pinhole collimation (Fig.2a), each beam is determined by two small pinholes, one in the source aperture with diameter of $2r_1$ and the other in the sample aperture with diameter of $2r_2$. For the multi-slit collimation (Fig.2b), each beam is determined by two small slits, i.e., one in the source aperture with width of $d_1$ and height of $h_1$ and the other in the sample aperture with width of $d_2$ and height of $h_2$. Apertures between the source and sample apertures are used to eliminate cross-talk between channels. The geometry of each small pinhole or slit channel is just the same as traditional pinhole SANS with $L_{SSD}=L_{SDD}=L$, $r_1=2r_2$, $d_1=2d_2$ and $h_1=2h_2$, but slit width and pinhole size are usually one order of magnitude smaller than traditional pinhole ones.

Beam current at the sample site of traditional pinhole geometry has been deduced [26] as Eq. (3). Beam current at the sample site of multi-pinhole VSANS can be written analogously as

$$I_{Multi-pinhole} = n_{mp} \frac{d^2\phi_s}{d\Omega d\lambda} \Delta\lambda \left(\frac{\pi}{4}\right) A_{2mp} \left(\frac{Q_{min}}{k}\right)^2 \quad (6)$$

where $n_{mp}$ means the number of pinholes in each multi-pinhole aperture and $A_{2mp}$ is the area of a single small pinhole in the sample aperture.

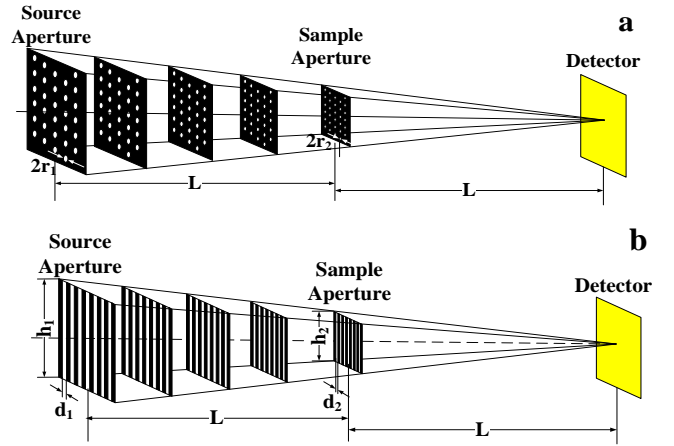

Fig. 2. Schematic view of the geometry of a. multi-pinhole and b. multi-slit apertures.

For multi-slit VSANS

$$Q_{min\,ms} = k\theta_{min} = k\frac{D_D}{2L} = k\frac{d_1}{L} \quad (7)$$

where $\theta_{min}$ is the same as the pinhole geometry shown in Fig. 1a and $D_D$ is the width of direct beam at the detector. Using Eq. (2) and Eq. (7), the beam current of multi-slit VSANS can be reduced as

$$I_{Multi-slit} = n_{ms} \frac{d^2\phi_s}{d\Omega d\lambda} \Delta\lambda \frac{d_1 h_1 d_2 h_2}{L^2} =$$

$$n_{ms} \frac{d^2\phi_s}{d\Omega d\lambda} \Delta\lambda {h_2}^2 \left(\frac{Q_{min}}{k}\right)^2 \quad (8)$$

where $n_{ms}$ is the number of narrow slits.

With Eq. (6)/Eq. (3) and Eq. (8)/Eq. (3), the gain of multi-pinhole VSANS and multi-slit VSANS over traditional pinhole SANS can be written as

$$\frac{I_{Multi-pinhole}}{I_P} = n_{mp}\frac{A_{2mp}}{A_{2P}} = n_{mp}\frac{r_2^2}{R_{2P}^2} \quad (9)$$

$$\frac{I_{Multi-slit}}{I_P} = n_{ms}\frac{h_2^2}{(\frac{\pi}{4})A_{2P}} = n_{ms}\frac{4h_2^2}{\pi^2 R_{2P}^2} \quad (10)$$

Considerable neutron beam current gain can be estimated for multi-slit VSANS over pinhole SANS at the same $Q_{min}$ according to Eq. (10), since the height of the sample slit $h_2$ is one order of magnitude greater than the diameter of the sample aperture $2R_{2P}$. Multi-slit VSANS gives up resolution in the vertical direction to get tremendous beam current gain, but its data has to be desmeared properly to get the same reduced data as pinhole SANS.

In practice, two factors have to be taken into consideration in designing multi-slit or multi-pinhole VSANS, i.e. gravity and parasitic scattering (scattering from the edge of the apertures or off-specula reflection from the side wall of collimators or reflecting mirrors). The gravity selective effect on the multi-pinhole option makes it unsuitable for TOF instruments, but there is no such constraint on the multi-slit. Parasitic scattering is another key factor to be addressed. It is a common phenomenon in all kinds of collimators, including pinhole collimators and soller collimators [29] [32], resulting in background around the beam stop, but its mechanism has not been fully understood [25]. There are several ways to mitigate parasitic scattering. First, it is effective to reduce the surface that induces parasitic scattering. For instance, multi-slit collimators have much fewer scattering surfaces than soller collimators. Second, one can polish the surfaces and edges that cause parasitic scattering. Third, one can use materials with large absorption cross-section or with negative scattering length ($^{113}$Cd, for example) [31] which are easy to process or polish.

## 3.3 Compound Refractive Lenses (CRL)

According to quantum mechanics, neutrons can also be treated like waves, and can therefore be refracted when passing through the interface of two materials with different scattering length densities. Unlike visible light, the refractive index of neutron is less than one for most materials, and is of the magnitude of $10^{-5}$ to $10^{-6}$. In 1998, Eskildsen et al [41] proposed and demonstrated multiple biconcave lenses or Compound Refractive Lenses (CRLS) to extend the minimum Q of conventional SANS instruments. Nowadays, reactor-based SANS equipped with focusing lenses can be found all over the world [10, 11, 42], and work has been done to analyze its effect to improve the performance of SANS [24, 43].

For a biconcave lens, the focal length, $f_o$, is given as

$$f_o = \frac{R}{2(1-n)} = \frac{R}{\delta} = \left(\frac{R}{\rho b_c}\right)\left(\frac{\pi}{\lambda^2}\right) \qquad (11)$$

where, R is the radius of curvature of the biconcave lens, ρ is the atomic density, $b_c$ is the bound average coherent scattering length of atoms, and λ is the wavelength of incident neutrons. When N thin biconcave lenses are used in series, the focal length is given as $f = f_o/N$. Note that the focal length depends on $\lambda^{-2}$. Therefore, neutrons with well-defined wavelength distribution are required for good focusing. Most SANS instruments at reactors use neutrons with a wavelength spread σ=Δλ/λ= 10 - 15 %, thus chromatic aberration should be expected.

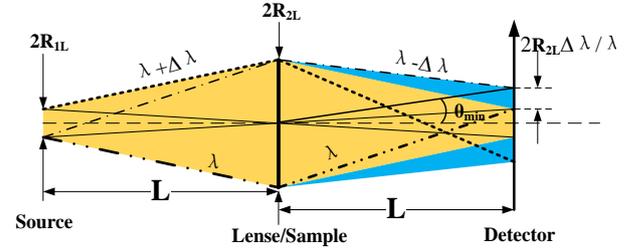

Fig. 3 Focusing geometry using ideal lenses. (Subscript L denotes Lens)

Glinka et al. have analyzed the possible gain of lenses over pinhole collimation [25] without considering chromatic aberration. But in the realm of VSANS, chromatic aberration of the lenses has to be considered. As shown in Fig. 3, the minimum Q of an ideal lens without thickness can be rendered as

$$Q_{\min} = k\theta_{\min} = k\frac{R_{1L}+2\sigma R_{2L}}{L} = k\frac{R_{1L}(1+2\sigma t)}{L} \qquad (12)$$

where $2\sigma R_{2L}$ is the aberration of the focal spot with σ=Δλ/λ [44] and $t=R_{2L}/R_{1L}$. Using Eq. (2) and Eq. (12), neutron current at the sample site for the lens system can be written as

$$I_L = \frac{d^2\phi_s}{d\Omega d\lambda}\Delta\lambda A_{2L}\left(\frac{Q_{min}}{k(1+2\sigma t)}\right)^2 T_L \qquad (13)$$

where $T_L$ is the transmission of the lenses.

Combining Eq. (1) and Eq. (12),

$$2R_{1P} = R_{1L}(1+2\sigma t) \qquad (14)$$

Eliminating $R_{1L}$ with Eq. (14) and setting $t=R_{2L}/R_{1L}$,

$$t = \frac{R_{2L}/R_{1P}}{2-2\sigma R_{2L}/R_{1P}} \qquad (15)$$

We can then combine Eq. (3) and Eqs. (13) and (15) and eliminate $Q_{min}/k$:

$$\text{Gain} = \frac{I_L}{I_P} = \frac{\pi A_{2L}T_L}{\left(\frac{\pi}{4}\right)A_{2P}(1+2\sigma t)^2} = \frac{16R_{2L}^2 T_L}{R_{1P}^2(1+2\sigma t)^2} =$$

$$16\left(\frac{R_{2L}}{R_{1P}}\right)^2 T_L\left(1-\sigma\frac{R_{2L}}{R_{1P}}\right)^2 \qquad (16)$$

Eq. (16) shows us a parabolic curve with variable $R_{2L}/R_{1P}$ and determined by two parameters: transmission $T_L$ and wavelength distribution σ=Δλ/λ. The peak point of the curve is at $R_{2L}/R_{1P}=1/(2\sigma)$. With Δλ/λ range from 10% to 15%, the gain of CRLS can be expected to be $100T_L$-$25T_L$ according to Eq. (16). However, CRLS cannot be easily applied to TOF instruments where chromatic aberration is inevitable.

### 3.4 Sextupole magnetic lenses

A neutron trajectory can be slightly bent when going through an inhomogeneous magnetic field, due to the translator forces acting on the magnetic dipole moments [45]. Thus, unpolarized neutrons can be polarized by a quadrupole magnet [46], and polarized neutrons can be focused or defocused by a sextuople magnet [47]. The concept of using a sextupole magnet as a neutron focusing lens was proposed by Farago in 1964 [45], but it was not developed until 1997, when Oku and Shimizu et al [47-50] started to use this technique and applied it to SANS-J-II [51] in Japan.

Two kinds of magnetic sextupole magnet can be used for neutron focusing, i.e., permanent sextupole magnets [47] and electromagnet sextupoles [52]. The former is more compact and maintenance free, the latter is not compact and has a low magnetic field, but with the help of superconducting sextupole magnets, neutron beams with large cross-sections can be focused [50].

The focal condition of the sextupole magnet is determined by the magnet length and neutron wavelength λ approximately as follows [53]

$$Z_f = \frac{2}{3}Z_m + \frac{1}{Z_m G \alpha}\left(\frac{h}{m_n \lambda}\right)^2 \qquad (17)$$

where $Z_f$ is the focal length, $Z_m$ is the magnet length, h is Planck's constant, G is the coefficient of the magnetic field gradient, $\alpha = |\mu_n/m_n| = 5.77 m^2 s^{-2} T^{-1}$, and $\mu_n$ is the neutron magnetic moment. According to Eq. (17), $Z_f$ is proportion to $\lambda^{-2}$, so the same chromatic aberration occurs in sextupole magnetic lenses just as in CRLS. The focusing geometry of the magnetic lenses is the same as that of material lenses (Fig. 3), as is beam current gain in Eq. (16) with transmission $T_L=1$.

Efforts have been made to develop permanent sextupole magnets for pulsed neutron sources by using a triplet permanent magnetic lens system [54] or two-nested permanent sextupole ring structure called modulating Permanent-Magnet Sextupole (mod-PMSx) [53]. Both devices can synchronize with the pulsed source to focus a wide bandwidth of neutrons. Mod-PMSx devices with a very compact design have been reported focusing a wide bandwidth of neutrons from 27–55 Å, and there remains a possibility of focusing cold neutrons from 3 to 15 Å [16]. According to Eq. (16), with no material absorption ($T_L=1$) and no chromatic aberration (σ=Δλ/λ=0), the gain of such mod-PMSx devices could be written as

$$\text{Gain} = \frac{I_{\text{Mod}-\text{PMSx}}}{I_P} = 16(R_{2L}/R_{1P})^2 \qquad (18)$$

### 3.5 Focusing mirrors: toroidal Wolter mirror

Grazing incidence reflection optics like ellipsoid mirrors and Wolter mirrors are able to focus neutrons without chromatic aberration. The technique has long been considered as the most promising means of focusing neutrons for SANS.

The Wolter mirror technique was first developed in X-ray telescopes for astronomy [55]. Efforts were made to use this technique in neutron focusing but failed [56] because of severe parasitic scattering. In recent years, with the developments of grinding, polishing and electroforming, perfect flat mirror surfaces with 2-4 Å rms roughness have become feasible [17, 57]. Khaykovich at Massachusetts Institute of Technology (MIT) has tested a four-nested Nickel Wolter mirror at the MIT Reactor [57], and pointed out its applicability to compact neutron sources [58] and high resolution neutron imaging [59]; Mildner at NIST Center for Neutron Research (NCNR) tested a type I Wolter optic at the NG7 SANS instrument for neutron imaging and tomography [60]. Here we will analyze the focusing geometry of a Type-I Wolter mirror based on the B.S. thesis of Bagdasarova at MIT [61] and calculate the possible gain over pinhole SANS.

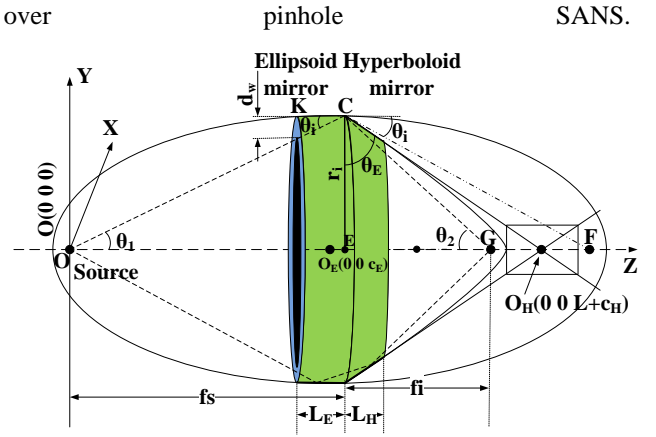

Fig. 4. Optical geometry of Type-I Wolter focusing mirror.

The geometry of the toroidal Wolter mirror is shown schematically in Fig. 4. Neutrons are emitted from the origin point or left focal point of the ellipsoid mirror O, reflected twice by the ellipsoid mirror and the hyperboloid mirror respectively and then reach point G which is the left focal point of the hyperboloid mirror.

The optical geometry of the Type-I Wolter mirror obeys the Abbe sine condition [62]

$$\frac{\sin\theta_1}{\sin\theta_2} = \frac{f_i}{f_s} = M \quad (19)$$

where $\theta_1$ and $\theta_2$ are the angles between the optical axis and incident plus reflected rays. M is the magnification. Given that $\theta_1$ and $\theta_2$ are very small, the Abbe sine condition can be fulfilled over the entire mirror and a perfect scaled image of the source can be formed at the focal point without coma.

The geometry of the Type-I Wolter focusing mirror can be defined by three parameters: L (distance from source to focal point), $\theta_i$ (grazing incident angle from the inner surface of ellipsoid and hyperboloid mirror) and M (magnification). All the other parameters $f_s$, $f_i$, $\theta_1$, $\theta_2$ (as shown in Fig. 4), $r_i$ (radius of the mirror), $a_E$, $b_E$, $c_E$ (long axis, short axis and focal length of the ellipsoid), $a_H$, $b_H$, $c_H$ (long axis, short axis and focal length of the hyperboloid) can be represented by these three parameters. According to Eq. (19) and the geometry of reflection in Fig. 4

$$\begin{cases} M = \frac{f_i}{f_s} \\ f_i + f_s = L \\ M = \frac{\theta_1}{\theta_2} \\ \theta_i = \frac{1}{4}(\theta_1 + \theta_2) \end{cases} \quad (20)$$

Solving the equations in Eq. (20) gives

$$\begin{cases} f_s = \frac{L}{1+M} \\ f_i = \frac{ML}{1+M} \\ \theta_1 = \frac{4\theta_i M}{1+M} \\ \theta_2 = \frac{4\theta_i}{1+M} \end{cases} \quad (23)$$

We define angle $\angle ECF = \theta_E$, and get

$$\theta_E = 180° - 2\theta_i - (90° - \theta_1) = 90° + 2\theta_i - \theta_2 \quad (24)$$

$$EF = r_i \tan(\theta_E) = \frac{r_i}{\tan(\theta_2 - 2\theta_i)} \quad (25)$$

With normal triangle CEO and known value of EF (Eq. (25)) we then have the following relations

$$\begin{cases} r_i = f_s \tan(\theta_1) \\ c_E = \frac{1}{2}\left(\frac{r_i}{\tan(\theta_2 - 2\theta_i)} + f_s\right) \\ c_H = \frac{1}{2}\left(\frac{r_i}{\tan(\theta_2 - 2\theta_i)} - f_i\right) \end{cases} \quad (26)$$

In the Descartes coordinate of Fig. 4 in the YZ plane, the ellipse and hyperbolic formula can be written as

$$\begin{cases} \frac{(z-c_E)^2}{a_E^2} + \frac{y^2}{b_E^2} = 1 \\ \frac{(z-L-c_H)^2}{a_H^2} - \frac{y^2}{b_H^2} = 1 \end{cases} \quad (27)$$

Substituting the formulas in Eq. (27) with point C (0, $r_i$, $f_s$) and together with the relations of the short and long axis of the ellipse and hyperbolic curve,

$$\begin{cases} \frac{(f_s-c_E)^2}{a_E^2} + \frac{r_i^2}{b_E^2} = 1 \\ \frac{(f_s-L-c_H)^2}{a_H^2} - \frac{r_i^2}{b_H^2} = 1 \\ c_E^2 = a_E^2 - b_E^2 \\ c_H^2 = a_H^2 + b_H^2 \end{cases} \quad (28)$$

Eq. (28) can be solved by numerical operation of Matlab software.

According to Eq. (1)

$$Q_{\min} = k\theta_{\min} = k\frac{R_s M}{f_i} = k\frac{R_s f_i}{f_s f_i} = k\frac{R_s}{f_s} \quad (29)$$

where $R_s$ is the radius of the source aperture.

Combining Eq. (2) and Eq. (29), the neutron current at the sample site can be written as

$$I_w = \frac{d^2\phi_s}{d\Omega d\lambda}\Delta\lambda\pi A_{2w}\left(\frac{Q_{min}}{k}\right)^2 \quad (30)$$

Dividing Eq. (30) by Eq. (3) gives

$$\text{Gain} = \frac{I_w}{I_P} = \frac{\pi A_{2w}}{\left(\frac{\pi}{4}\right)A_{2P}} = \frac{4A_{2w}}{A_{2P}} = \frac{4\pi(r_i^2 - (r_i - d_w)^2)}{\pi R_{2P}^2} \quad (31)$$

where $A_{2w}$ is the annular area between the circular neutron absorption diaphragm and the ellipsoid mirror, and $d_w$ is the width of this annular belt. The coordinate of point K can be determined by substituting the ellipse formulas in Eq. (27) with point K (0, y, $f_s$-$L_E$) (Eq. (32)) and $d_w$ can be calculated accordingly (Eq. (33)).

$$y = b_E * \text{sqrt}\left(1 - \frac{(f_s-L_E-c_E)^2}{a_E^2}\right) \quad (32)$$

$$d_w = b_E * \text{sqrt}\left(1 - \frac{(f_s-L_E-c_E)^2}{a_E^2}\right) - \frac{r_i(f_s-L_E)}{f_s} \quad (33)$$

### 3.6 Other techniques

Novel multi-beam VSANS was proposed at ILL [63] based on tests with D33. With very small source and sample aperture, multiple beams come from multiple reflection from the neutron guide, and reach multiple points at the detector. It is a unique technique that has considerable gain over traditional pinhole SANS with very small samples. The technique can only be used in certain configurations, however, and the overlap of the scattering patterns cannot be easily deconvoluted.

### 3.7 Comparison of various VSANS techniques

In order to get a quantitative impression of all these VSANS techniques, a comparison of the beam current gain has to be made at Q values of 0.002 Å$^{-1}$ and 0.0002 Å$^{-1}$. We assume a typical reactor-based neutron source with selector ($\lambda$=6.28Å, $\Delta\lambda/\lambda$=10%) and maximum source aperture of 60×60 mm. Pinhole geometry, as shown in Fig. 1 a. Pinhole geometry; b. Extremely long pinhole geometry (Subscript 1 denotes source aperutre, 2 denotes sample aperutre, P denotes Pinhole, E denotes Extremely long, respectively). , will be set as a bench mark for all other techniques, with $2R_{1P}$ =30 mm, $2R_{2P}$=15 mm, L=15 m and $Q_{min}$= k*$2R_{1P}$/L =0.002 Å$^{-1}$.

The radius of the CRLS and magnetic lenses are set to be 15 mm. The length of the extremely long instrument is assumed to be nL=4L=60 m. The geometry of Wolter mirror VSANS is asymmetric and $f_s$ cannot be equal to $f_i$. We assume $f_s+f_i$=30 m, M=0.6, $\theta_i$=0.01 rad=0.57°, $L_m=L_E$=0.5 m. With Eq. (23) (26) (28) and (33), the parameters of the Wolter mirror can be calculated to be: $f_s$=18.75 m, $f_i$=11.25 m, $r_i$=0.28 m and $d_w$=0.005 m.

The gain factor and gains of the VSANS techniques at $Q_{min}$=0.002 Å$^{-1}$ and $Q_{min}$=0.0002 Å$^{-1}$ over pinhole SANS are listed in Table 2. Note that all the gains are relative to the bench mark of pinhole geometry at $Q_{min}$=0.002 A$^{-1}$, so gains at $Q_{min}$=0.0002 Å$^{-1}$ have all been divided by 10000 according to Eq. (1) and Eq. (2). As shown in Table 1, only multi-slit aperture, mod-PMSx and Wolter mirror have gains greater than 0.1 at 0.0002 Å$^{-1}$ which means these three techniques are potential candidates of VSANS in the aspect of neutron beam current. However, advanced techniques are needed to produce the sophisticated and extremely precise devices of mod-PMSx and Wolter mirrors.

Table 1 Gain factor and beam current gain of VSANS techniques at $Q_{min}$=0.002 Å$^{-1}$ and $Q_{min}$=0.0002Å$^{-1}$

| VSANS technique | Gain factor | Gain at $Q_{min}$=0.002 Å$^{-1}$ | Gain at $Q_{min}$=0.0002 2Å$^{-1}$ |
|---|---|---|---|
| Pinhole geometry | -- | 1 | 0.0001 |
| Extremely long pinhole geometry | $\frac{R_{2E}^2}{R_{2P}^2}$ | 8 ($R_{2E}$=2$R_{2P}$) | 0.0016 ($R_{2E}$=4$R_{2P}$) |
| Multi-pinhole | $n_{mp}\frac{r_2^2}{R_{2P}^2}$ | 4 ($n_{mp}$=4, $r_2/R_{2P}$=1) | 0.01 ($n_{mp}$=100, $r_2/R_{2P}$=1) |
| Multi-slit | $n_{ms}\frac{4h_2^2}{\pi^2 R_{2P}^2}$ | 13 ($n_{ms}$=2, $h_2/R_{2P}$=4) | 0.65 ($n_{ms}$=10, $h_2/R_{2P}$=40) |
| Material Lenses | $16\left(\frac{R_{2L}}{R_{1P}}\right)^2 T_L \left(1-\sigma\frac{R_{2L}}{R_{1P}}\right)^2$ | 10.4 ($R_{2L}/R_{1P}$=1, $\sigma$=0.1, $T_L$=0.8) | 0.008 ($R_{2L}/R_{1P}$=5, $\sigma$=0.1, $T_L$=0.8) |
| Magnetic lenses | $16\left(\frac{R_{2L}}{R_{1P}}\right)^2\left(1-\sigma\frac{R_{2L}}{R_{1P}}\right)^2$ | 13 ($R_{2L}/R_{1P}$=1, $\sigma$=0.1) | 0.01 ($R_{2L}/R_{1P}$=5, $\sigma$=0.1) |
| mod-PMSx (synchronized with pulsed source) | $16\left(\frac{R_{2L}}{R_{1P}}\right)^2$ | 16 ($R_{2L}/R_{1P}$=1) | 0.16 ($R_{2L}/R_{1P}$=10) |
| Wolter mirror | $\frac{r_i d_w - d_w^2}{\frac{1}{4}R_{2P}^2}$ | 198 ($r_i$=0.28m, $d_w$=0.005m, $R_{2P}$=7.5mm) | 1.98 ($r_i$=0.28m, $d_w$=0.005m, $R_{2P}$=0.75mm) |

## 4 Conceptual design of multi-slit VSANS for CSNS

Based on the above analysis, we propose a multi-slit Very Small Angle Scattering instrument for the second phase of the CSNS project. As shown in Fig. 5, the total length of the instrument is designed to be 40 m. Its features include a two-meter-long bender to avoid the direct-line-of-sight, swappable multi-slit and guide system to switch between pinhole SANS and VSANS mode, polarizing option and three detectors.

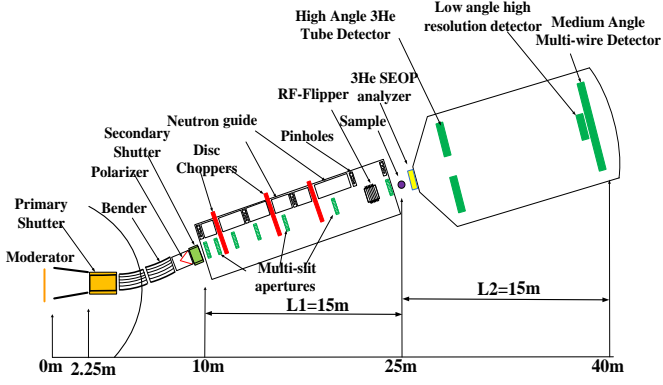

Fig. 5 Schematic view of the VSANS design for CSNS.

The bender design of the VSANS (JOINT) is similar to that of the Multi-purpose Reflectometer (MR) of CSNS [64]. The multi-slit aperture design follows the algorithm from Ref[65]. Neutrons are polarized by a V-shape polarizing supermirror and analyzed by $^3$He Spin Exchange Optical Pump (SEOP) absorber. Neutrons are detected by a multi-wire chamber (15×15 cm) or a scintillator detector with 1-2 mm resolution (VSANS mode) and $^3$He tube (100×100 cm) with 5-8 mm resolution (pinhole SANS at intermedium and high scattering angles). The design parameters of CSNS VSANS are summarized in Table 2.

Table 2 Design parameters of CSNS VSANS

| | |
|---|---|
| **Instrument Name** | VSANS (JOINT) |
| **Moderator** | Coupled Liquid Hydrogen |
| **Length** | 40 m |
| **L1** | 15 m |
| **L2** | 15 m |
| **Guide cross-section** | 40x40 mm |
| **Sample Size** | 20x20 mm |
| **Q range** | 0.0002-1 Å$^{-1}$ |
| **Wavelength range** | 2-18 Å |
| **Bender** | 2 m length, 100 m radius of curvature, m=3.6 supermirror coating |
| **Choppers** | Three choppers |
| **Collimation** | Multi-slit apertures Pinhole apertures |
| **Polarizer** | m=4 Fe/Si Supermirror |
| **Analyzer** | SEOP $^3$He absorber |
| **Detector** | Multi-wire Chamber 20×20 cm with 1-2 mm resolution $^3$He tube 100×100 cm with 5-8 mm resolution |

The performance of the CSNS VSANS pinhole and multi-slit VSANS modes have been evaluated and compared by using the Mcstas simulation package with the parameters listed in Table 2. To simplify the geometry and reduce calculation time, the bender and choppers were not included in the simulations. The multi-slit converging aperture system has 7 slits in each aperture with $d_1 = 2d_2 = 3$ mm and $h_1 = 2h_2 = 40$ mm, while pinhole mode has $R_1=15$ mm and $R_2=7.5$ mm. Simulated distribution of direct neutron intensity distribution at the detector of multi-slit VSANS mode and pinhole SANS mode are displayed in Fig. 6. Full Width at Half Maximum (FWHM) of horizontal distribution of mulit-slit and pinhole mode are 3.1 mm and 27.4 mm respectively. With Eq. (10), the gain of multi-slit VSANS over pinhole SANS can be calculated to be

$$\frac{I_{multi-slit}}{I_{Pinhole}} = \frac{n_{ms}\frac{4h_2^2}{\pi^2 R_{2P}^2}}{10000} = \frac{7\frac{4*20^2}{\pi^2 0.75^2}}{10000} = 0.202 \quad (34)$$

Here, we choose $R_{2P}$ equals 0.75 mm, which is ten times smaller than $R_2$=7.5 mm, to fulfill the condition of Eq. (10), which requires $Q_{min}$ of multi-slit and pinhole be the same. Therefore, the result of Eq. (10) is divided by 10000 to get the right gain of the multi-slit mode. The calculated gain factor in Eq. (34) is similar to the McStas simulation result, 6.08E5/3.03E6=0.201, as shown in Fig. 6.

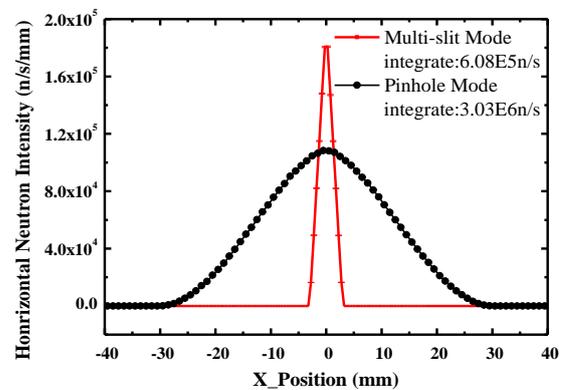

Fig. 6 Horizontal neutron intensity distribution at the detector for mulit-slit and pinhole mode.

# 5 Conclusion

Different techniques of Very Small Angle Neutron Scattering (VSANS) have been reviewed and beam current gains over traditional pinhole SANS have been analyzed and quantitatively compared. Considerable gains over traditional pinhole SANS are expected. Each technique has its own features and is applicable to specific source and geometric conditions. Based on the analysis and comparison, a conceptual design for a multi-slit VSANS instrument (JOINT) is proposed for the China Spallation Neutron Source. It will be the first VSANS instrument in the world based at a spallation neutron source, if funding can be allocated.

# Acknowledgement

We are grateful to Dr. Peter Willendrup (DTU) for his help using McStas, and the Software Group of CSNS for valuable discussions and simulation environment. We also thank Dr. Sylvain Desert (LLB), Dr. John Barker (NCNR) and Dr. Daniel Clemens (HZB) for their discussion about the techniques of multi-slit and multi-pinhole VSANS.

# References

1. http://www.nobelprize.org/nobel_prizes/physics/laureates/1994/, retrieved 5th November 2015
2. L. A. Feigin and D. I. Svergun, *Structure Analysis by Small Angle X-ray and Neutron Scattering*. (New York: Springer. 1987), p. 339
3. W. Marshall and S. W. Lovesey, *Theory of ThermalNeutron Scattering*. (New York: Oxford University Press. 1971), p. 600
4. J. Schelten, Kerntechnik, **14:** 86-88 (1972)
5. K. Ibel, Journal of Applied Crystallography, **9:**296-309 (1976)
6. J. FItter;, T. Gutberlet; and J. Katsaras, *Neutron Scattering in Biology Techniques and Applications*. (Berlin Heidelberg New York: Springer. 2006),
7. D. T. Bowron, A. K. Soper, K. Jones, et al., Rev Sci Instrum, **81:**033905 (2010)
8. R. K. Heenan, J. PENFOLD and S. M. KING, J. Appl. Cryst., **30:**7 (1997)
9. C. J. Glinka, J. G. Barker, B. Hammouda, et al., Journal of Applied Crystallography, **31:**430-445 (1998)
10. E. P. Gilbert, J. C. Schulz and T. J. Noakes, Physica B: Condensed Matter, **385-386:**1180-1182 (2006)
11. Y.-S. Han, S.-M. Choi, T.-H. Kim, et al., Nuclear Instruments and Methods in Physics Research Section A: Accelerators, Spectrometers, Detectors and Associated Equipment, **721:**17-20 (2013)
12. A. Radulescu, V. Pipich, H. Frielinghaus, et al., Journal of Physics: Conference Series, **351:**012026 (2012)
13. P. Lindner and R. Schweins, Neutron News, **21:**15-18 (2010)
14. V. T. Sylvain Desert, Julian Oberdisse, Annie Brulet, J. Appl. Cryst., **40:**471-477 (2007)
15. T. Oku, H. Iwase, T. Shinohara, et al., Journal of Applied Crystallography **40:**408-413 (2007)
16. M. Yamada, Y. Iwashita, M. Ichikawa, et al., Progress of Theoretical and Experimental Physics, **2015:**43G01-40 (2015)
17. J. Guo, S. Takeda, S. Y. Morita, et al., Opt Express, **22:**24666-24677 (2014)
18. D. Liu, B. Khaykovich, M. V. Gubarev, et al., NatCommun, **4:**2556 (2013)
19. K. Lefmann and K. Nielsen, Neutron News, **10:**20-23 (1999)
20. W. Yin, T. J. Liang and Q. Z. Yu, Nuclear Instruments and Methods in Physics Research Section A:Accelerators, Spectrometers, Detectors and Associated Equipment, **631:**105-110 (2011)
21. K. Lieutenant, P. Lindner and R. Gahler, Journal of Applied Crystallography, **40:**1056-1063 (2007)
22. R. Cubitt, R. Schweins and P. Lindner, Nuclear Instruments and Methods in Physics Research Section A: Accelerators, Spectrometers, Detectors and Associated Equipment, **665:**7-10 (2011)
23. D. T. Mildner and J. Carpenter, J. Appl. Cryst.,

**17:**249-256 (1984)
24. K. C. Littrell, Nucl. Instr. Meth. A, **529:**22-27 (2004)
25. C. J. Glinka, J. G. Barker and D. F. R. Mildner, Nuclear Instruments and Methods in Physics Research Section A: Accelerators, Spectrometers, Detectors and Associated Equipment, **795:**122-127 (2015)
26. S. M. Choi, J. Barker, C. J. Glinka, et al., J. Appl. Cryst., **33:**793-796 (2000)
27. *http://www.ncnr.nist.gov/programs/sans/pdf/Seminar_2014_11_19_JBarker.pdf, retrieved 5th November 2015*
28. *http://www.ansto.gov.au/ResearchHub/Bragg/Facilities/Instruments/Bilby/index.htm, retrieved 5th November 2015*
29. A. C. Nunes, Nuclear Instruments and Methods, **119:**291-293 (1974)
30. M. Carpenter and J. Faber, Jnr, Journal of AppliedCrystallography, **11:**464-465 (1978)
31. C. J. Glinka, J. M. Rowe and J. G. LaRock, Journal of Applied Crystallography, **19:**427-439 (1986)
32. P. Thiyagarajan, J. E. Epperson, R. K. Crawford, et al., Journal of Applied Crystallography, **30:**280-293 (1997)
33. F. M. A. Margaça, A. N. Falcão, J. F. Salgado, etal., Physica B: Condensed Matter, **276-278:**189-191 (2000)
34. A. N. Falcao, F. M. A. Margaca and F. G. Carvalho, Applied Physics A: Materials Science & Processing, **74:**s1462-s1464 (2002)
35. A. N. Falcão, F. M. A. Margaça and F. G. Carvalho, Journal of Applied Crystallography, **36:**1266-1269 (2003)
36. A. Len, G. Pépy and L. Rosta, Physica B: Condensed Matter, **350:**E771-E773 (2004)
37. K. Vogtt, M. Siebenbürger, D. Clemens, et al., Journal of Applied Crystallography, **47:**237-244 (2014)
38. F. Klose, P. Constantine, S. J. Kennedy, et al., Journal of Physics: Conference Series, **528:**012026 (2014)
39. S. Jaksch, D. Martin-Rodriguez, A. Ostermann, et al., Nuclear Instruments and Methods in Physics Research Section A: Accelerators, Spectrometers, Detectors and Associated Equipment, **762:**22-30 (2014)
40. https://europeanspallationsource.se/, retrieved 5th November 2015,
41. M. R. Eskildsen, Nature, **391:**563-566 (1998)
42. S. Okabe, T. Karino, M. Nagao, et al., Nuclear Instruments and Methods in Physics Research Section A: Accelerators, Spectrometers, Detectors and Associated Equipment, **572:**853-858 (2007)
43. B. Hammouda and D. F. Mildner, J. Appl. Cryst.,**40:**250-259 (2007)
44. F. P. Doty, J. T. Cremer, M. A. Piestrup, et al., **5541:**75-114 (2004)
45. P. S. Farago, Nuclear Instruments and Methods, **30:**271-273 (1964)
46. K. Taketani, K. Mishima, T. Ino, et al., Physica B: Condensed Matter, **404:**2643-2645 (2009)
47. H. M. Shimizu, H. Kato, T. Oku, et al., Physica B: Condensed Matter, **241–243:**172-174 (1997)
48. H. M. Shimizu, Y. Suda, T. Oku, et al., Nuclear Instruments and Methods in Physics Research Section A: Accelerators, Spectrometers, Detectors and Associated Equipment, **430:**423-434 (1999)
49. T. Oku, J. Suzuki, H. Sasao, et al., Nuclear Instruments and Methods in Physics Research Section A: Accelerators, Spectrometers, Detectors and Associated Equipment, **529:**116-119 (2004)
50. J. Suzuki, T. Oku, T. Adachi, et al., Journal of Applied Crystallography, **36:**795-799 (2003)
51. S. Koizumi, H. Iwase, J.-i. Suzuki, et al., Physica B: Condensed Matter, **385-386:**1000-1006 (2006)
52. A. Steinhof, Nuclear Instruments and Methods in Physics Research Section A: Accelerators, Spectrometers, Detectors and Associated Equipment, **397:**371-379 (1997)
53. M. Yamada, Y. Iwashita, M. Ichikawa, et al., Physica B: Condensed Matter, **404:**2646-2651 (2009)
54. T. Oku, H. Kira, T. Shinohara, et al., Journal of Physics: Conference Series, **251:**012078 (2010)
55. http://imagine.gsfc.nasa.gov/science/toolbox/xray_telescopes2.html, retrieved 5th November 2015
56. B. Alefeld, C. Hayes, F. Mezei, et al., Physica B, **234-236:**1052-1054 (1997)
57. B. Khaykovich, M. V. Gubarev, Y. Bagdasarova, etal., Nuclear Instruments and Methods in Physics Research Section A: Accelerators, Spectrometers, Detectors and Associated Equipment, **631:**98-104 (2011)
58. B. Khaykovich, M. V. Gubarev, V. E. Zavlin, et al., Physics Procedia, **26:**299-308 (2012)
59. D. Liu, D. Hussey, M. V. Gubarev, et al., Applied

Physics Letters, **102:**183508 (2013)

60. D. F. R. Mildner and M. V. Gubarev, Nuclear Instruments and Methods in Physics Research Section A: Accelerators, Spectrometers, Detectors and Associated Equipment, **634:**S7-S11 (2011)

61. Y. S. Bagdasarova, *Wolter Mirror Microscope: Novel Neutron Focussing and Imaging Optic*, B. S. Thesis. (Massachusetts: Massachusetts Institute of Technology, MIT. 2010)

62. R. Kingslake, *Lens Design Fundamentals*. (New York London Academic Press. 1978), p. 357

63. C. D. Dewhurst, Journal of Applied Crystallography, **47:**1180-1189 (2014)

64. F. Wang, T. Liang, W. Yin, et al., Science China Physics, Mechanics and Astronomy, **56:**2410-2424 (2013)

65. *http://www.ncnr.nist.gov/staff/hammouda/the_SANS_toolbox.pdf, retrieved 5th November 2015*